\documentclass[10pt, conference]{IEEEtran}
\IEEEoverridecommandlockouts
\usepackage{cite}
\usepackage{amsmath,amssymb,amsfonts}
\usepackage{algorithmic}
\usepackage{graphicx}
\usepackage{textcomp}
\usepackage{xcolor}
\usepackage[a4paper, total={184mm,239mm}]{geometry}
\usepackage{multirow}
\usepackage{threeparttable}
\usepackage{array}
\newcolumntype{P}[1]{>{\centering\arraybackslash}p{#1}}

\def\BibTeX{{\rm B\kern-.05em{\sc i\kern-.025em b}\kern-.08em
    T\kern-.1667em\lower.7ex\hbox{E}\kern-.125emX}}
\begin{document}

\title{SAF: Scalable Acceleration Framework for dynamic and flexible scaling of FPGAs}

\author{Masudul Quraishi, Michael Riera, Fengbo Ren, Aman Arora, Aviral Shrivastava}
% \IEEEauthorblockA{\textit{dept. name of organization (of Aff.)} \\
% \textit{name of organization (of Aff.)}\\
% City, Country \\
% email address or ORCID}
% \and
% \IEEEauthorblockN{2\textsuperscript{nd} Given Name Surname}
% \IEEEauthorblockA{\textit{dept. name of organization (of Aff.)} \\
% \textit{name of organization (of Aff.)}\\
% City, Country \\
% email address or ORCID}
% \and
% \IEEEauthorblockN{3\textsuperscript{rd} Given Name Surname}
% \IEEEauthorblockA{\textit{dept. name of organization (of Aff.)} \\
% \textit{name of organization (of Aff.)}\\
% City, Country \\
% email address or ORCID}
% \and
% \IEEEauthorblockN{4\textsuperscript{th} Given Name Surname}
% \IEEEauthorblockA{\textit{dept. name of organization (of Aff.)} \\
% \textit{name of organization (of Aff.)}\\
% City, Country \\
% email address or ORCID}
% \and
% \IEEEauthorblockN{5\textsuperscript{th} Given Name Surname}
% \IEEEauthorblockA{\textit{dept. name of organization (of Aff.)} \\
% \textit{name of organization (of Aff.)}\\
% City, Country \\
% email address or ORCID}
% \and
% \IEEEauthorblockN{6\textsuperscript{th} Given Name Surname}
% \IEEEauthorblockA{\textit{dept. name of organization (of Aff.)} \\
% \textit{name of organization (of Aff.)}\\
% City, Country \\
% email address or ORCID}
% }

\maketitle

\begin{abstract}
FPGAs are increasingly gaining traction in cloud and edge computing environments due to their hardware flexibility, low latency, and low energy consumption. However, the existing hardware stack of FPGA and the host-FPGA connectivity does not allow flexible scaling and simultaneous reconfiguration of multiple devices, which limits the adoption of FPGA at scale. In this paper, we present SAF -- an Ethernet-based scalable acceleration framework that allows FPGA to be hot-plugged into a network in a stand-alone fashion without connecting to a local host CPU, which enables flexible scalability. SAF provides a custom FPGA shell and a set of Ethernet protocols that allow FPGAs to connect with a remote host to accelerate application kernels. SAF can configure multiple FPGAs simultaneously, which significantly reduces the reconfiguration time in scaling effort. We implemented the SAF framework using Intel FPGA SDK for OpenCL and 20 Bittware 385A cards with Arria-10 FPGAs. We analyze a case study and conduct experiments to compare SAF with state-of-the-art multi-FPGA clusters. Results show that SAF provides 13X faster reconfiguration than sequential PCIe programming, reduces the hardware setup costs by 38\%, application runtime by 25\%, and energy consumption by 27\%. We evaluated the performance scalability of SAF using the PTRANS benchmark of the HPCC FPGA benchmark suite and showed an almost linear speedup for strong and weak scaling scenarios.
\end{abstract}


\section{Introduction}
Field Programmable Gate Arrays (FPGAs) are suitable candidates in both cloud and edge computing environments due to their hardware flexibility, low latency, and low energy consumption \cite{biookaghazadeh2018fpgas, xu2020case, chen2014enabling}. FPGAs are efficient in processing streaming data from input/output (I/O) at the network edge, and they can also provide consistently high computational throughput for accelerating both high-concurrency and high-dependency algorithms, serving a much broader range of cloud and edge applications \cite{xu2020case, leeser2021fpgas}. Even though commercial cloud providers, including Amazon, \cite{amazonf1} Microsoft \cite{catapult2014}, and Alibaba \cite{Alibaba} have integrated FPGAs into their services, the existing hardware stack of FPGA and the host-FPGA connectivity does not allow flexible scaling and simultaneous reconfiguration of multiple devices, which limits the adoption of FPGA at scale.

There are existing works on Ethernet-based host-FPGA communication that implement a network stack (Ethernet, TCP-IP, ARP, UDP) on FPGA for a single FPGA configuration and communication \cite{johansson2021fpga, shreejith2017vega, kucharczyk2013simple}. There are multi-FPGA works \cite{huang2009scalable, reggiani2021enhancing, mondigo2017design, liu2010building, sano2023essper, meyer2023multi} that accelerate a certain application. They use the PCIe-based flow for reconfiguration and execution. There are works on virtualization \cite{VirtioHWSB2018, bandara2024performance, zha2020virtualizing, fahmy2015virtualized} and multi-tenancy \cite{yamakura2021multi, mbongue2021performance} that reduce vendor-specific driver dependency and improve usability. However, the multi-FPGA reconfiguration support and flexible scalability for FPGA are missing. 

The existing multi-FPGA clusters \cite{sano2023essper, meyer2023multi} provide high throughput and low latency by using high-speed host-FPGA and inter-FPGA networks. However, these systems are not suitable for flexibly integrating new FPGAs. The inter-FPGA networks cannot be changed dynamically during execution. Furthermore, there is no hot plug capability to plug in FPGA without changing the host code or application. Even though they are using high-speed networks, which provide superior performance, they use the traditional PCIe flow for reconfiguration of FPGAs. PCIe flow and existing toolchain do not support the reconfiguration for multiple FPGAs. Furthermore, PCIe requires a local host CPU for reconfiguration and operation of the FPGAs, making scaling inconvenient and expensive. 

We propose \textbf{SAF}, an Ethernet-based \textbf{S}calable \textbf{A}cceleration \textbf{F}ramework that allows FPGAs to be hot-plugged into a network in a stand-alone fashion without connecting to a local host CPU, which enables flexible scalability. SAF provides a custom FPGA shell and a set of Ethernet protocols that allow FPGAs to connect with a remote host to accelerate application kernels. The FPGA shell is modified to route the Ethernet payload data through different interfaces to comply with the protocols. The standalone accelerator protocols presented in the paper allow (i) automatic network discovery of FPGAs, (ii) partial reconfiguration by the remote host, (iii) FPGA memory management, (iv) sending control commands to execute kernels, and (v) sending output results to host. SAF can configure multiple FPGAs simultaneously, which reduces the reconfiguration time while scaling. The SAF custom shell is developed using HDL and HLS flow of Intel FPGA SDK for OpenCL, which provides flexibility in developmental effort. 

The contribution of the paper is summarized as follows:
\begin{enumerate}
    \item We propose SAF, an Ethernet-based scalable acceleration framework for dynamic and flexible scalability of FPGAs. We have developed a custom FPGA shell and a set of accelerator protocols that allow FPGAs to connect and communicate to a remote host in a standalone fashion without the need for a local host. The remote host can configure the FPGA and run application kernels using the standalone accelerator protocols.
    \item We propose automatic network discovery of FPGA, enabling hot plug operation for FPGA without installing any driver. The hot plug capability enables seamless dynamic integration and flexible scalability of FPGAs. 
    \item SAF can reconfigure multiple FPGAs on the network simultaneously. This reduces the reconfiguration time while scaling up compared to the sequential PCIe-based reconfiguration. 
\end{enumerate}

We measure the reconfiguration time for two PCIe setups (two PCIe devices per host and a PCIe device tree (DT) hosted by a single host) and compare it with SAF. SAF provides 2x to 13x faster reconfiguration than PCIe and PCIe-DT devices. 

SAF can connect to a network in a stand-alone fashion without a local host. This reduces the cost of scaling compared to PCIe-based connectivity, which needs a local host. SAF reduces setup cost for scaling by up to 38\% as compared to SOTA multi-FPGA clusters Noctua \cite{meyer2023multi} and ESSPER \cite{sano2023essper}.

We evaluate the performance scalability of SAF using the PTRANS benchmark from the HPCC FPGA Benchmark Suite \cite{HPCC-FPGA} on up to 20 Bittware 385A FPGA Accelerator Cards \cite{Bittware}. We measure speedup and scaled speedup in strong and weak scaling scenarios to show that adding more FPGAs to the framework enables almost linear performance scalability. 
 
To evaluate the flexible scalability of SAF, we analyze a case study where a multi-FPGA cluster needs to scale to accommodate increasing on-demand computation. We compare the FPGA application runtime and energy consumption in an on-demand scaling scenario with two other state-of-the-art (SOTA) multi-FPGA clusters, Noctua \cite{meyer2023multi} and ESSPER \cite{sano2023essper}, to show that SAF can reduce the application runtime by 25\% and FPGA energy consumption by 27\%. 

\section{Related Work}
\subsection{Ethernet protocol and FPGAs}
Ethernet-based communication between host CPU and FPGA and remotely configuring FPGAs over Ethernet has been proposed in the literature \cite{kammerling2007fpga, alachiotis2010efficient, perrett2011simple, lieber2011fpga}. Most of the prior works implement a network stack (UDP, ARP, TCP/IP) on the FPGA \cite{johansson2021fpga, shreejith2017vega, kucharczyk2013simple} to communicate with the host. Most of the time, the application running on FPGA is limited to embedded use on a single FPGA \cite{xie2020achieving, shreejith2017vega, wang2023design}, and the scalability factor is not considered. 

\subsection{Scalability using Virtualization and Multi-tenancy}
VirtIO-based virtual machine for FPGA \cite{VirtioHWSB2018, bandara2024performance} provides an alternative to vendor-provided device-specific drivers. While this solves the PCIe driver dependency by providing a portable driver, the RTL development effort is significant and does not provide reconfiguration capability or scalability. Virtualization of multiple FPGAs in the cloud \cite{zha2020virtualizing, fahmy2015virtualized} is proposed, but the approaches use PCIe flow for connection and, therefore, lack flexible scalability. Research on multi-tenancy FPGAs \cite{yamakura2021multi, mbongue2021performance, dessouky2021sok} provides the sharing of single FPGA resources between multiple users, a form of scalability. However, multi-FPGA scaling for multi-tenancy is still very limited. 

\subsection{Heterogeneous FPGA Clusters and Multi-FPGA Systems}
Networked and heterogeneous FPGA clusters \cite{tarafdar2017enabling, tsoi2010axel, porrmann2010raptor} are proposed for cloud and edge computing. There are existing works on scalable FPGA architecture \cite{huang2009scalable, reggiani2021enhancing, mondigo2017design, liu2010building, choi2014map, arthanto2022fshmem, alonso2021elastic, hong2022dfx} primarily focus on accelerating applications, running emulation on multiple FPGAs and comparing the performance and power with other accelerators like GPU. Some of the multi-FPGA systems \cite{sano2023essper, meyer2023multi} connect FPGAs and host CPUs into a hybrid network. The FPGAs are also interconnected in a full duplex, point-to-point fashion. Even though these architectures provide increased performance and low latency, integrating new FPGAs is not straightforward. The inter-FPGA networks are programmed once and cannot be changed during execution. The heterogeneous clusters and multi-FPGA systems do not have hotplug support for flexible scalability. They use PCIe for reconfiguration and cannot configure multiple FPGAs simultaneously. 

% Table \ref{tab:qual_comparison} shows a comparison between SAF and existing Multi-FPGA clusters Noctua and ESSPER.
% \begin{table}[!hbt]
% \scriptsize
% \centering
% \caption{\label{tab:qual_comparison}Features and Network architecture comparison between SAF and existing Multi-FPGA clusters Noctua and ESSPER}
% \begin{tabular}{|p{0.7cm}|p{0.6cm}|p{1.2cm}|p{0.5cm}|p{0.4cm}|p{0.7cm}|p{0.7cm}|c|}
% \hline
% \textbf{\begin{tabular}[c]{@{}c@{}}Multi-\\ FPGA \\ system\end{tabular}} & \textbf{\begin{tabular}[c]{@{}c@{}}Perf. \\ Scaling\end{tabular}} & \textbf{\begin{tabular}[c]{@{}l@{}}Setup \\ cost\end{tabular}} & \textbf{\begin{tabular}[c]{@{}c@{}}Multi \\ FPGA \\ prog.\end{tabular}} & \textbf{\begin{tabular}[c]{@{}c@{}}Hot\\plug\end{tabular}} & \textbf{\begin{tabular}[c]{@{}c@{}}Inter \\ FPGA \\ network\end{tabular}} & \textbf{\begin{tabular}[c]{@{}c@{}}CPU\\FPGA \\ Network\end{tabular}} & \textbf{\begin{tabular}[c]{@{}c@{}}Network \\ between \\ CPUs\end{tabular}} \\ \hline
% Noctua & Yes & \begin{tabular}[c]{@{}l@{}}Host CPUs\\ Increase\\cost.\end{tabular} & No & No & Yes & Yes & Yes \\ \hline
% ESSPER & Yes & \begin{tabular}[c]{@{}l@{}}Host CPUs \\ Increase\\cost\end{tabular} & No & No & Yes & Yes & Yes \\ \hline
% SAF & Yes & \begin{tabular}[c]{@{}l@{}}No additional \\ host CPU \\reduces cost\end{tabular} & Yes & Yes & No & Yes & No \\\hline
% \end{tabular}
% \end{table}

\section{State-of-the-Art Multi-FPGA Clusters vs. SAF}
\label{section3-system-architecture}
\begin{figure}[t]
  \includegraphics[width=3.6in]{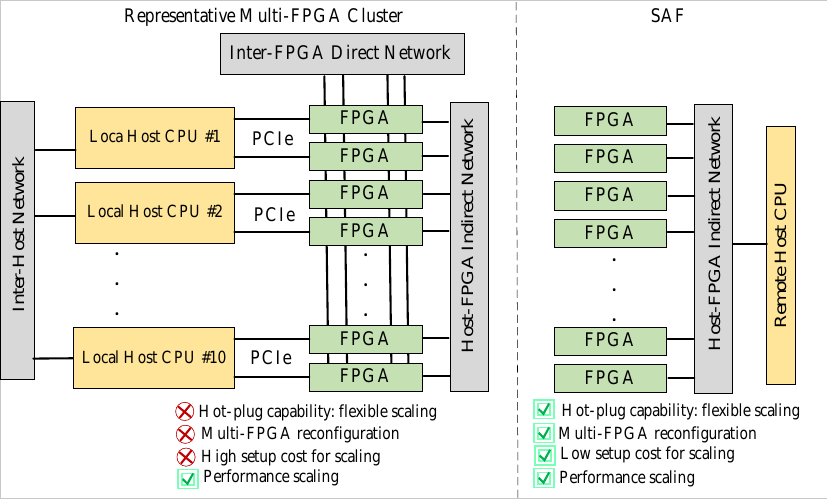}
  \centering
  \caption{The system overview and benefits of SAF (right) compared to a representative SOTA multi-FPGA cluster (Left). The cluster uses a hybrid of host-FPGA indirect network and FPGA-FPGA direct network, which provides superior performance but lacks flexibility in scaling. SAF only uses an indirect network with a remote host and provides hot plug integration of FPGAs, which enables flexible scalability.}
  \label{fig:overview}
\end{figure}%

% \avs{You need to justify this SOTA architecture by referring to previous works.}

Figure \ref{fig:overview} shows the system overview of SAF and compares it with a representation of the multi-FPGA clusters Noctua \cite{meyer2023multi} and ESSPER \cite{sano2023essper}. Each local host CPU is connected to two FPGAs in the multi-FPGA cluster using PCIe. It has a hybrid of host-FPGA indirect network and FPGA-FPGA direct network. The host CPUs are also interconnected via a network. The inter-FPGA direct network allows different connection configurations depending on how it is programmed. While this setup is excellent for low latency communication and high throughput, it is not easy to integrate a new FPGA into the system. The inter-FPGA network is programmed once before execution and cannot be changed dynamically. The FPGAs are tied to local hosts in a PCIe-based flow, which does not allow the reconfiguration of multiple FPGAs simultaneously. Furthermore, new local host CPUs are needed while scaling up, which increases setup effort and cost for scaling.  

In SAF, FPGAs and the remote host CPU are connected in an indirect Ethernet network. There is no inter-FPGA direct network. The FPGAs can be hot-plugged into the network without affecting the current execution of the remote host application and FPGA kernels. This enables dynamic and flexible scaling of FPGAs. In SAF, the remote host can configure multiple FPGAs simultaneously over the network, which reduces the reconfiguration time of multiple FPGAs compared to the PCIe-based flow. The setup cost for scaling is lower compared to the multi-FPGA clusters due to a single remote host.
 
\section{SAF Architecture}
\begin{figure}[!hbt]
  \includegraphics[width=3.5in]{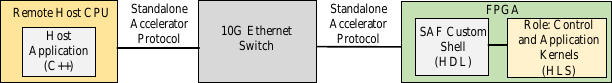}
  \centering
  \caption{The high-level architecture of SAF showing the key components of the framework. The SAF custom shell and the kernels in the role are designed in such a way that they can communicate with the remote host application using standalone accelerator protocols.}
  \label{fig:components}
\end{figure}

There are four key components in the SAF architecture: (i) SAF custom shell, (ii) Control and application kernels, (iii) Remote host application (iv) Standalone accelerator protocols. Figure \ref{fig:components} shows the high level connectivity between the components. 
% The FPGAs and the remote host CPU communicate using the Ethernet Protocols described in section \ref{section-SAF}. The FPGAs can be hot-plugged into the system by physically connecting them to the switch. Multiple network switches can be cascaded to increase the number of FPGAs in the network. As soon as the FPGAs are connected, the remote host can identify them using a network discovery packet containing the FPGA's unique MAC address. The remote host runs an application that keeps track of the available devices on the network, creates and sends an Ethernet payload to configure the device and manage the device kernels, collecting application results from the devices and organizing and storing the results. The host can have one or multiple Ethernet ports that connect to the network switch. The host CPU has a unique MAC address, which the devices use to communicate using Ethernet protocols. 

\subsection{SAF Custom Shell}
\label{section-shell}
\begin{figure}[!hbt]
  \includegraphics[width=3.3in]{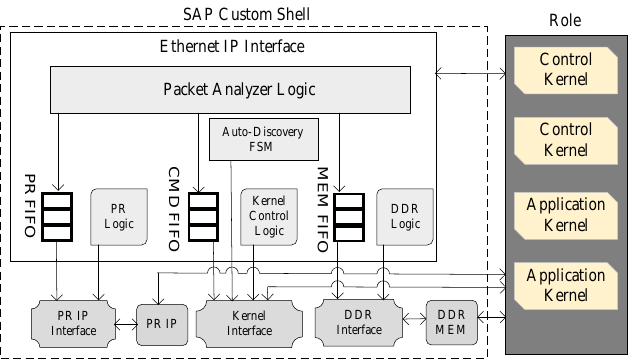}
  \centering
  \caption{The micro-architecture of SAF custom shell showing the interconnection between module interfaces, logic blocks, and kernels. The SAF custom shell analyzes and routes the payload data from the Ethernet packets to appropriate interfaces. The control kernels complement the custom shell to enable compliance with the standalone accelerator protocols.}
  \label{fig:shell}
\end{figure}

We have developed a custom shell for SAF so that the FPGAs can comply with the standalone accelerator protocols.  The SAF custom shell is developed by modifying the default MAC-type shell that comes with the Bittware 385A accelerator cards \cite{Bittware}. The default shell receives data from the Intel OpenCL host code via PCIe and routes them to different module interfaces. The purpose of the custom shell is to read Ethernet packets received from the remote host, analyze the packets, and route the data to these modules. To achieve this, the SAF custom shell modifies four key IP interfaces from the default shell. (i) The Ethernet IP interface, (ii) The partial reconfiguration (PR) IP interface, (iii) The kernel interface, and (iv) The DDR interface. The four interfaces use Intel's Avalon Memory-Mapped (MM) interface \cite{Avalon} and can communicate via an address-based read/write of host-agent connections. Figure \ref{fig:shell} shows the interconnection between the interfaces and custom logic blocks in the SAF custom shell. 

\subsubsection{Ethernet IP Interface}
The Ethernet IP is responsible for the exchange of Ethernet packets between the remote host and FPGAs. Ethernet packets at the IP interface are received by the Avalon streaming (ST) \cite{Avalon} inputs. The IP interface HDL is modified to add the following four logics: packet analyzer logic, auto-discovery finite state machine (FSM), PR logic, kernel control logic, and DDR logic.

The packet analyzer logic extracts the packet header and stores the data into three different FIFOs depending on the packet type. The packet types 0x80AA, 0x80CC, and 0x80DD indicate PR bitstream, kernel control, and DDR memory data, and they are stored in PR FIFO, CMD FIFO, and MEM FIFO, respectively. An asynchronous FIFO module from Intel is instantiated to implement the FIFOs. The data from these FIFOs are sent to the output of the Ethernet IP interface, which is routed to PR IP, kernel interface, and DDR interface via Avalon MM host-agent write logic. Since the different interfaces operate under different clocks, the Avalon clock crossing bridge module is inserted between the MM connections. The asynchronous FIFOs and clock crossing bridges ensure a safe clock transition between the modules.

The auto-discovery FSM sends a kernel execution command to the kernel interface when it detects the first network packet after the FPGA is plugged into the network. This command launches the control kernel (discussed in section \ref{section: kernel}) responsible for sending the discovery packet to the remote host. The PR logic, kernel control logic, and DDR logic added to the Ethernet IP interface are responsible for reconfiguration, kernel execution, and memory management. The logic blocks are discussed in subsequent subsections. 

\subsubsection{Partial Reconfiguration IP Interface}
The PR IP is the primary logic responsible for the reconfiguration of FPGA.  In PCIe flow, the OpenCL host function sends the bitstream data to the PCIe IP, which routes it to the PR IP. In SAF, the host application sends the bitstream data to the Ethernet IP. Therefore, we need a channel to send the bitstream data from the Ethernet IP to the PR IP. We implemented a wrapper logic around the PR IP interface that multiplexes bitstream data between the PCIe IP and the Ethernet IP to achieve this. The default selection for the mux is PCIe. Whenever the Ethernet IP receives bitstream data, the multiplexer selects the data from the Ethernet, given that the PCIe programming channel is not currently occupied. The PR logic added to the Ethernet IP interface controls the write and read of bitstream data from the PR FIFO. Once the programming via Ethernet is completed, a done signal is asserted. This signal is used to send a PR confirmation from FPGA to the remote host. 

\subsubsection{Kernel Interface}
In the PCIe flow, the host sends the kernel execution commands using API functions, which send the control data to the kernel interface via PCIe. In SAF, the remote host executes the kernel by sending Ethernet packets. The payload of the kernel control packet consists of an address and data. The address and data can differ depending on the kernel's order in the kernel pipeline. As discussed in section III, a separate FIFO is added to the Ethernet IP to store the kernel command data (see Figure \ref{fig:shell}). The Ethernet IP output and the kernel interface's input are connected using the Avalon MM interface. Kernel control logic is added to the Ethernet IP interface to read the kernel command data from the CMD FIFO and send it to the kernel interface. 

\subsubsection{DDR Interface}
In the PCIe flow, the host enqueues memory buffers containing input data to the FPGA DDR memory using PCIe. The kernel interface reads the input data from a specific address of the memory. In SAF, the input data from the Ethernet packet is stored in MEM FIFO in the Ethernet IP (see Figure \ref{fig:shell}). The data width for the DDR interface is 512 bits, where the data in the Ethernet packet is stored as 64 bits. Therefore, DDR logic is added to pop the data from the MEM FIFO and convert the 64-bit packets into 512 bits. Then, DDR logic stores the data in a specific address in DDR memory, depending on the order of the argument in the kernel. After reading the input data, the kernel interface sends an acknowledgment to the remote host. 

\subsection{Control and Application Kernels}
\label{section: kernel}
The control and application kernels are configured in the reconfigurable region (role) of the FPGA (see Figure \ref{fig:shell}). The shell logic is implemented using HDL. On the other hand, the control and application kernels are developed in OpenCL using the high-level synthesis (HLS) flow. The mix of HDL and HLS flow provides flexibility in the SAF framework design. 

The control kernels complement the SAF custom shell by sending information about FPGA, creating Ethernet packets, and sending the results back to the remote host. For example, when the FPGA is connected to the network, a control kernel sends a discovery packet to send information about the FPGA to the host and ensure the FPGA is discoverable in the network. Another control kernel helps to create Ethernet packets using the output data from the application kernel and sends them back to the host. The application kernel is the application that is accelerated on FPGA. The control kernels, application kernels, and shell are compiled together to create the bitstream file, which is used to configure the FPGA.

\subsection{Remote Host Application}
In PCIe flow, the host application manages the application kernel using OpenCL runtime and PCIe drivers. The OpenCL host includes APIs to control the platform, manage memory, and execute programs on the FPGA. In SAF, we implement the host code using C++; no additional drivers are needed. The host application uses socket programming to directly access the Ethernet port on the CPU for sending and receiving Ethernet packets. The host is responsible for detecting and managing FPGAs on the network. The host application generates Ethernet packets following the standalone accelerator protocol. The data for the Ethernet payload is read from a file on the host machine. For example, while reconfiguration of FPGAs, the data is read from the raw bitstream file (.rbf) to generate Ethernet packets. The host application also stores and displays the output result from the application kernel.

\begin{table}[!hbt]
\scriptsize
\centering
\caption{\label{tab:packet} Different communication protocols supported by SAF}
\begin{tabular}{|p{1.6cm}|l|p{3.1cm}|l|}
\hline
\textbf{Protocol} & \textbf{\begin{tabular}[c]{@{}l@{}}Packet \\ Type\end{tabular}} & \textbf{Payload Data} & \textbf{Communication} \\ \hline
\begin{tabular}[c]{@{}l@{}}Auto Discovery\end{tabular} & 0x80EF & \begin{tabular}[c]{@{}l@{}}Device MAC address, \\ vendor and product IDs\end{tabular} & FPGA -\textgreater Host \\ \hline
\begin{tabular}[c]{@{}l@{}}Partial \\ Reconfiguration\end{tabular} & 0x80AA & \begin{tabular}[c]{@{}l@{}}Bitstream data \\ for reconfiguration\end{tabular} & Host -\textgreater FPGA \\ \hline
\begin{tabular}[c]{@{}l@{}}PR Confirmation\end{tabular} & 0x80AB & \begin{tabular}[c]{@{}l@{}}Reconfiguration \\ Acknowledgement\end{tabular} & FPGA -\textgreater Host \\ \hline
\begin{tabular}[c]{@{}l@{}}Kernel input\end{tabular} & 0x80DD & \begin{tabular}[c]{@{}l@{}}Data for kernel argument \\ to be saved in DDR\end{tabular} & Host -\textgreater FPGA \\ \hline
\begin{tabular}[c]{@{}l@{}}Input data\\ confirmation\end{tabular} & 0x80DB & \begin{tabular}[c]{@{}l@{}}Input data read \\ acknowledgement\end{tabular} & FPGA -\textgreater Host \\ \hline
\begin{tabular}[c]{@{}l@{}}Kernel Execution\end{tabular} & 0x80CC & \begin{tabular}[c]{@{}l@{}}Kernel execution command: \\ Address and the command data\end{tabular} & Host -\textgreater FPGA \\ \hline
\begin{tabular}[c]{@{}l@{}}Output results\end{tabular} & 0x80CB & \begin{tabular}[c]{@{}l@{}}The output after \\ running application kernel\end{tabular} & FPGA -\textgreater Host \\ \hline
\end{tabular}

\end{table}
\begin{figure}[!hbt]
  \includegraphics[width=3.2in]
  {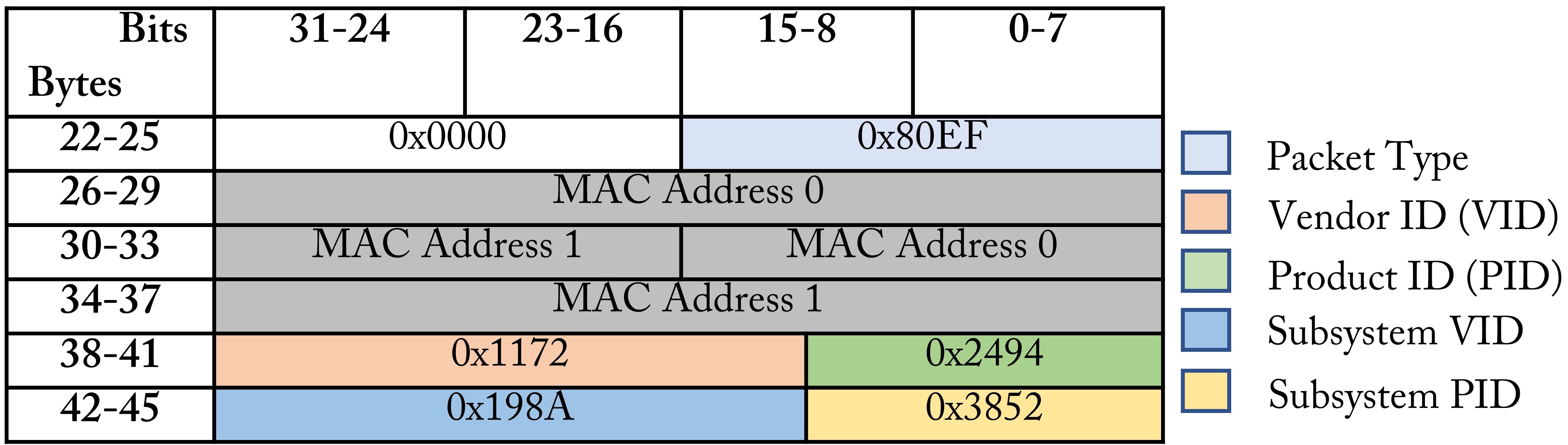}
  \centering
  \caption{Ethernet Auto-Discovery Packet sent by FPGAs to the remote host. Using this packet, the FPGAs can announce their presence on the network by sharing their unique MAC addresses, vendor IDs, and product IDs.}
  \label{fig:discovery}
\end{figure}%

\subsection{Standalone Accelerator Protocols}
\label{section-SAF}
The standalone accelerator protocols are the set of protocols that the remote host and FPGA need to comply with for SAF operations. We have identified five fundamental operations that must be supported by the SAF framework: (i) automatic network discovery of FPGAs, (ii) partial reconfiguration by the remote host, (iii) FPGA memory management, (iv) sending control commands to execute kernels, and (v) sending output results. We present a set of Ethernet protocols between the remote host and FPGA to support these operations in Table \ref{tab:packet}.

\begin{figure}[!b]
  \includegraphics[width=3.5in]{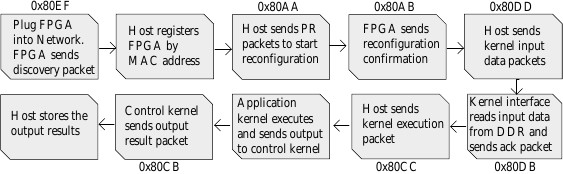}
  \centering
  \caption{Execution flow diagram showing the sequence of host-FPGA communications using the standalone accelerator protocols listed in Table \ref{tab:packet}. This flow gives an overview of how SAF enables application acceleration on FPGAs.}
  \label{fig:exe}
\end{figure}%

As an example, we show the automatic discovery packet in Figure \ref{fig:discovery}. In the PCIe flow, FPGA devices connected to the PCIe can be discovered using an API call from the host machine. In SAF, FPGA devices send network discovery packets to the host to announce their availability. The network discovery packets are sent automatically by an FPGA as soon as they are connected to the network switch. The discovery packet contains the unique MAC address to identify the device and general information like vendor ID and product ID of the device. For simplicity, we only show the Ethernet packet type and payload part of the packet. The preamble, source, and destination MAC, CRC bits, and padding bits \cite{ethernet} are not shown. The two MAC addresses are for the two MAC channels in 385A accelerator cards. We have used only MAC address 0 for SAF. 
The execution flow diagram in Figure \ref{fig:exe} shows the order of the five operations in SAF and the protocols used. 

\section{Experimental Setup}
To evaluate the flexible scalability, setup cost, performance scaling, and reconfiguration time of SAF, we used 20 Bittware 385A accelerator cards \cite{Bittware} with Intel Arria 10 FPGAs and a MAC board support package for 10G Ethernet connectivity. 

For the experiments using the PCIe flow, we connect the 20 FPGA boards to 10 edge host nodes using PCIe Gen3 x8 connections. The host machine on the edge nodes has an Intel Xeon Processor E3-1275 v5 with 8M Cache, a 3.60 GHz processor, and 32GB DDR4 memory. A USB-JTAG connection exists between the edge host and the FPGAs for the initial shell configuration. 

For the experiments with SAF, the FPGA cards and the Ethernet host machine are connected to the Ethernet network using two Dell X4012 network switches. Each network switch is equipped with 12 10 Gigabit SFP+ ports. The QSFP+ ports of the FPGAs are connected to the network switch using the Molex adapter and cable. We connected 10 FPGAs to one switch and 10 FPGAs and the host machine to another switch. The remote host machine has an Intel Xeon E5-2637 v3 CPU with a 3.5GHz processor, 15M Cache, and 64GB memory. 

We use the PTRANS benchmark from the HPCC FPGA benchmark suite \cite{HPCC-FPGA}. HPCC FPGA is an OpenCL-based benchmark suite with a focus on high-performance computing. The PTRANS benchmark computes the transpose of a quadratic matrix and saves the result in a memory buffer. 

In the case of SAF, we remotely configured 20 FPGAs simultaneously using the PTRANS bitstream by sending a broadcast packet. We then separately send the kernel argument data to each FPGA. While kernel execution, we can again send the kernel control packet using a broadcast packet and execute the kernels simultaneously. 

\section{Experiment Results}
% \avs{I think this experiment cannot be the first one... since this is a derived result. Maybe this should be the last experiment. I think the text in the experiments section can be tightened to fit in the page budget.}

\subsection{13X Faster Reconfiguration time}
We compare the reconfiguration time for SAF with the PCIe-based programming flow. For PCIe programming flow, we consider two different host-FPGA connectivity. First, each host has two FPGAs connected with PCIe, and the CPUs are connected via an Ethernet network. This architecture is similar to ESSPER. For multiple FPGA configurations in this architecture, the bitstream can be sent to the CPUs via the network, and then the host application can send the bitstream via PCIe. Second, all FPGAs are connected to a single host using a PCIe device tree (PCIe-DT). We reconfigure 1-20 FPGAs using the PTRANS bitstream and record the configuration time. We assume the initial shell configuration is done using USB-JTAG and we only measure the time to partial reconfiguration of PTRANS bitstream, keeping the shell unchanged.

\begin{figure}[!h]
  \includegraphics[width=3in]{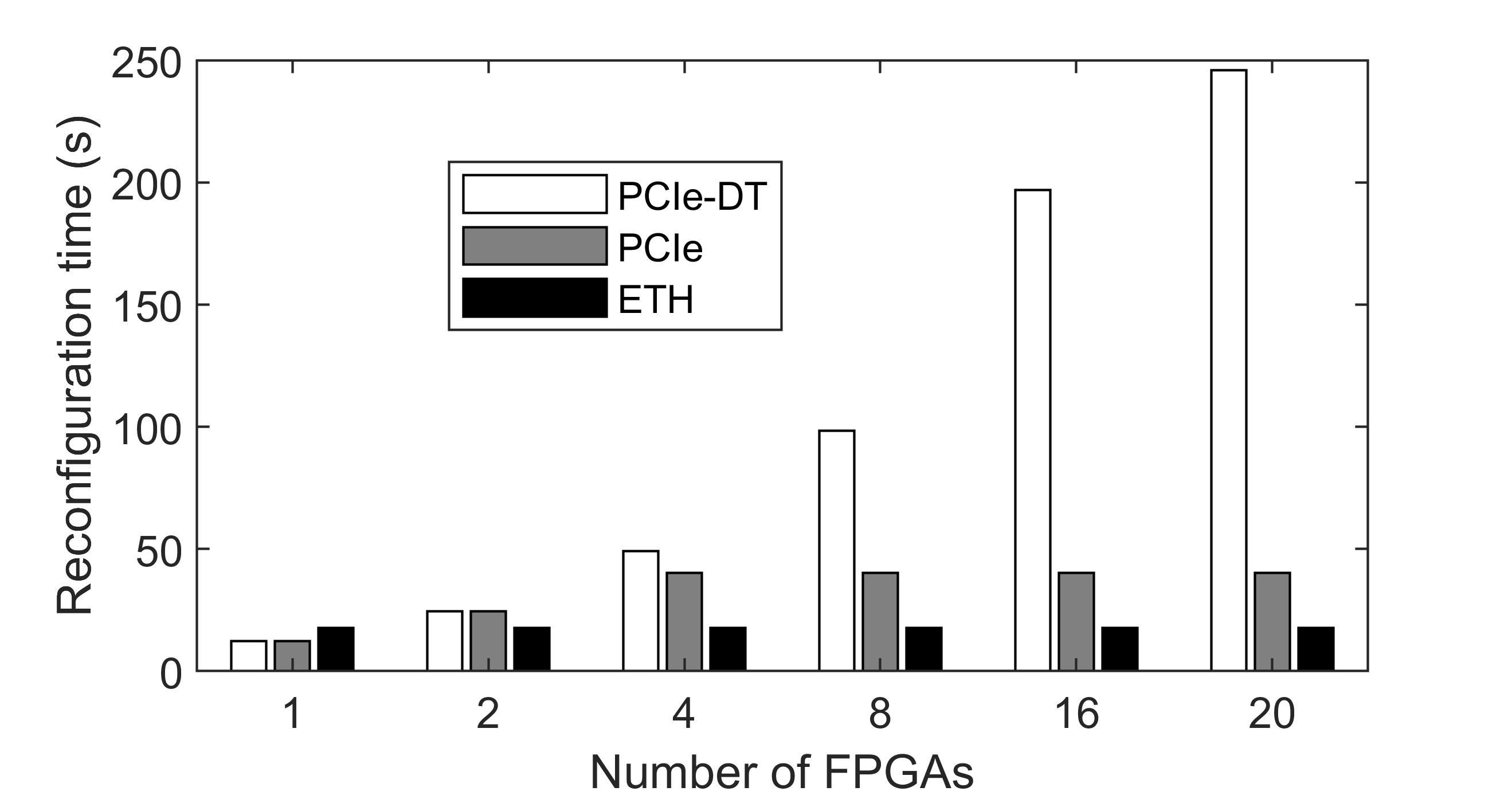}
  \centering
  \caption{Partial Reconfiguration time for PTRANS bitstream for configuring 1-20 FPGAs for PCIe, PCIe device tree (DT), and Ethernet (SAF). SAF can reconfigure multiple FPGAs simultaneously, which reduces reconfiguration time compared to PCIe and PCIe-DT.}
  \label{fig:reconfig}
\end{figure}%

Figure \ref{fig:reconfig} shows the partial reconfiguration time for SAF, PCIe, and PCIe-DT. The PTRANS bitstream file size is 97.4 MB. To program a single FPGA, PCIe and PCIe-DT take about 12.3 seconds, whereas Ethernet (ETH) takes 17.76 seconds. This is because the host overhead to create packets and transfer the bitstream data over Ethernet is greater than the OpenCL host code overhead to transfer via PCIe. For two FPGAs, it takes double the time (24.60 seconds) to reconfigure PCIe and PCIe-DT connected FPGAs since the programming is done sequentially by a single host. As we scale up, PCIe will have a reconfiguration time of 24.60 seconds, plus an additional 15.67 seconds to send the bitstream over the network. Since there is a separate host for every two FPGAs, they can be configured in parallel once the hosts have the bitstream. The programming time for ETH remains unchanged at 12.3 seconds since it uses the same broadcast packet to program the FPGAs simultaneously. For PCIe-DT, the reconfiguration time multiplies with the number of devices since a single host application needs to program them sequentially. For 20 FPGAs, the time to reconfigure PCIe-DT, PCIe, and ETH-connected devices are 246, 40.27, and 17.76 seconds, respectively. Therefore, SAF is able to provide 2.27x to 13.85x faster reconfiguration than PCIe and PCIe-DT devices. 

\begin{table}[!hbt]
\scriptsize
\centering
\caption{\label{tab:setup-cost} Setup cost comparison between multi-FPGA clusters and SAF}
\begin{tabular}{|p{0.5cm}|p{0.2cm}p{0.2cm}c|ccc|p{0.65cm}|}
\hline
\multirow{2}{*}{\textbf{FPGA}} & \multicolumn{3}{c|}{\textbf{\begin{tabular}[c]{@{}c@{}}Number of Hosts\end{tabular}}} & \multicolumn{3}{c|}{\textbf{\begin{tabular}[c]{@{}c@{}}Cost (USD)\end{tabular}}} & \multirow{2}{*}{\textbf{\begin{tabular}[c]{@{}c@{}}\%Cost \\ Savings\end{tabular}}} \\ \cline{2-7}
 & \multicolumn{1}{c|}{\textbf{Noc}} & \multicolumn{1}{c|}{\textbf{ESSP}} & \textbf{SAF} & \multicolumn{1}{c|}{\textbf{Noc}} & \multicolumn{1}{c|}{\textbf{ESSP}} & \textbf{SAF} &  \\ \hline
1 & \multicolumn{1}{c|}{1} & \multicolumn{1}{c|}{1} & 1 & \multicolumn{1}{c|}{1849.98} & \multicolumn{1}{c|}{1849.98} & 1849.98 & 0.00 \\ \hline
2 & \multicolumn{1}{c|}{2} & \multicolumn{1}{c|}{1} & 1 & \multicolumn{1}{c|}{3699.96} & \multicolumn{1}{c|}{2599.97} & 2599.97 & 0.00 \\ \hline
4 & \multicolumn{1}{c|}{4} & \multicolumn{1}{c|}{2} & 1 & \multicolumn{1}{c|}{7399.92} & \multicolumn{1}{c|}{5199.94} & 4099.95 & 21.15 \\ \hline
8 & \multicolumn{1}{c|}{8} & \multicolumn{1}{c|}{4} & 1 & \multicolumn{1}{c|}{14799.84} & \multicolumn{1}{c|}{10399.88} & 7099.91 & 31.73 \\ \hline
12 & \multicolumn{1}{c|}{12} & \multicolumn{1}{c|}{6} & 1 & \multicolumn{1}{c|}{22199.76} & \multicolumn{1}{c|}{15599.82} & 10099.87 & 35.26 \\ \hline
16 & \multicolumn{1}{c|}{16} & \multicolumn{1}{c|}{8} & 1 & \multicolumn{1}{c|}{29599.68} & \multicolumn{1}{c|}{20799.76} & 13099.83 & 37.02 \\ \hline
20 & \multicolumn{1}{c|}{20} & \multicolumn{1}{c|}{20} & 1 & \multicolumn{1}{c|}{36999.60} & \multicolumn{1}{c|}{25999.70} & 16099.79 & 38.08 \\ \hline
\end{tabular}
\end{table}

\subsection {21\% - 38\% Reduced Hardware Setup Cost for Scaling}
We compare the cost of setup for scaling up the Noctua, ESSPER, and SAF.  For fairness, we only compare the cost of CPUs and FPGAs and assume that the same CPUs and FPGAs are used in all the architectures. Table \ref{tab:setup-cost} shows the total setup cost for scaling up from one to twenty FPGAs. We can see that SAF can save 21.15\% - 38.08\% costs compared to Noctua and ESSPER. The minimum cost between Noctua and ESSPER is considered while calculating the savings. In Noctua and ESSPER, each host CPU connects to one or two FPGAs via PCIe. They do not use PCIe switches to connect more than two FPGAs to a single host. Therefore, for scaling up, the multi-FPGA clusters require additional host CPUs, which increases the cost of the hardware setup. 

\subsection{Almost linear Performance Scaling}
We run the PTRANS benchmark from the HPCC benchmark suite on the SAF framework to evaluate performance scaling. A matrix of 32,768 elements is transposed using strong and weak scaling on 20 FPGAs. In a strong scaling scenario, the number of FPGA is increased, keeping the matrix size the same. In a weak scaling scenario, the matrix size per FPGA remains the same. For both strong and weak scaling, We get an almost linear speedup as more FPGAs are added.

\begin{figure}[!hbt]
\centering
   \includegraphics[width=2.4in]{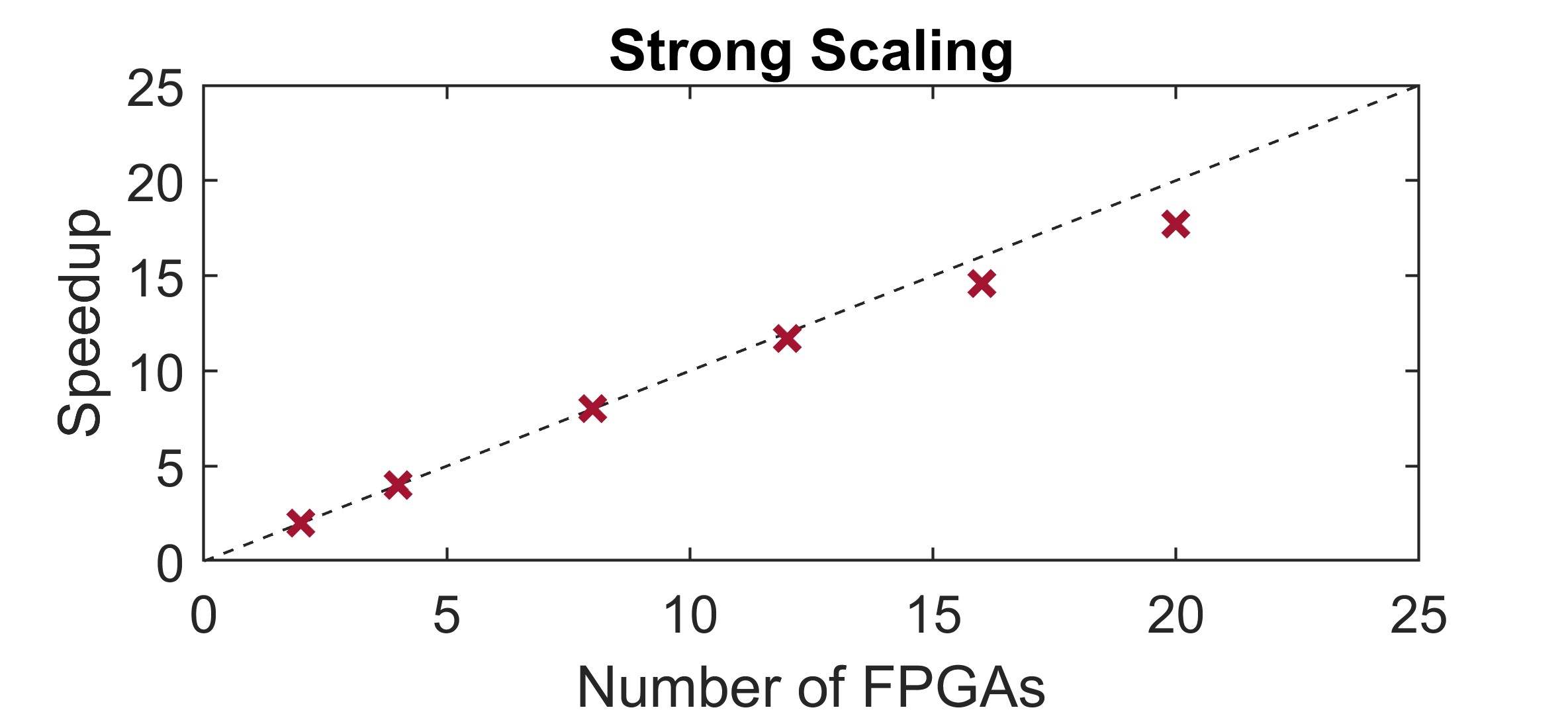}
   \caption{Speedup of the PTRANS benchmark executed on 20 FPGAs in a strong scaling scenario. At a higher number of FPGAs, due to host overhead and data transfer overhead, the scaling deviates from the optimal (black-dotted) line. }
   \label{fig:scaling}
 \end{figure}

In Figure \ref{fig:scaling},  the black dotted line is the ideal scaling behavior. We can see that when the number of FPGAs is 8 or less, the scaling behavior follows the optimal line. At a higher number of FPGAs, due to host overhead and data transfer overhead, the scaling deviates from the optimal line. 

\subsection{25\% Reduced Runtime and 27\% Reduced Energy}
To evaluate the flexible scalability of SAF, we analyze a case study of a simulation running on a multi-FPGA cluster. We consider the two multi-FPGA clusters Noctua \cite{meyer2023multi} and ESSPER \cite{sano2023essper} and SAF, having a baseline architecture with 4 FPGAs. The FPGAs are running an application (simulation of molecular structures) with a runtime of 10 hours. Let's assume, to reduce the application runtime we can scale the architecture to add 4 more FPGAs while the application is still running. The timeline to finish the application with the scaled architecture will depend on at what stage the scaling is done. 

Let's assume the scaling occurs after 4 hours when the simulation is 40\% completed. For Noctua and ESSPER, the current execution needs to stop to add more FPGAs, reprogram the inter-FPGA network and then execution restarts. After scaling, the clusters needed to complete 10 hours of simulation, but with double resources. So, the overall runtime will be 4+(10/2) = 9 hours. While the runtime is better than the baseline (10 hours), it loses the initial 4 hours. For SAF, four FPGAs can be hot-plugged dynamically into the network, and the host CPU can start using them. Since the execution is ongoing, after 4 hours, they need to complete 6 hours of simulation but with a scaled architecture. So total time taken by SAF will be 4+ (6/2) = 7 hours, which is 22.22\% less than the clusters. 

\begin{table}[!ht]
\centering
\caption{\label{tab:case-study-time} Application Runtime and Energy reduction by SAF}
\begin{tabular}{|P{1.1cm}|cc|cc|P{1cm}|P{1cm}|}
\hline
\multirow{2}{*}{\textbf{\begin{tabular}[c]{@{}c@{}}\% of \\ simulation\\ completed\end{tabular}}} & \multicolumn{2}{c|}{\textbf{Cluster}} & \multicolumn{2}{c|}{\textbf{SAF}} & \multirow{2}{*}{\textbf{\begin{tabular}[c]{@{}c@{}}\\\%Time \\ reduction\end{tabular}}} & \multirow{2}{*}{\textbf{\begin{tabular}[c]{@{}c@{}}\\\%Energy \\ reduction\end{tabular}}} \\ \cline{2-5}
 & \multicolumn{1}{c|}{\textbf{\begin{tabular}[c]{@{}c@{}}Time \\ (h)\end{tabular}}} & \textbf{\begin{tabular}[c]{@{}c@{}}Energy \\ (kJ)\end{tabular}} & \multicolumn{1}{c|}{\textbf{\begin{tabular}[c]{@{}c@{}}Time \\ (h)\end{tabular}}} & \textbf{\begin{tabular}[c]{@{}c@{}}Energy \\ (kJ)\end{tabular}} &  &  \\ \hline
0 & \multicolumn{1}{c|}{5} & 9.79 & \multicolumn{1}{c|}{5} & 9.79 & 0.00 & 0.00 \\ \hline
10 & \multicolumn{1}{c|}{6} & 10.77 & \multicolumn{1}{c|}{5.5} & 9.83 & 8.33 & 8.74 \\ \hline
20 & \multicolumn{1}{c|}{7} & 11.75 & \multicolumn{1}{c|}{6} & 9.86 & 14.29 & 16.05 \\ \hline
30 & \multicolumn{1}{c|}{8} & 12.73 & \multicolumn{1}{c|}{6.5} & 9.90 & 18.75 & 22.23 \\ \hline
40 & \multicolumn{1}{c|}{9} & 13.71 & \multicolumn{1}{c|}{7} & 9.94 & 22.22 & 27.53 \\ \hline
50 & \multicolumn{1}{c|}{10} & 9.79 & \multicolumn{1}{c|}{7.5} & 9.97 & 25.00 & -1.82 \\ \hline
60 & \multicolumn{1}{c|}{10} & 9.79 & \multicolumn{1}{c|}{8} & 10.01 & 20.00 & -2.23 \\ \hline
70 & \multicolumn{1}{c|}{10} & 9.79 & \multicolumn{1}{c|}{8.5} & 10.04 & 15.00 & -2.53 \\ \hline
80 & \multicolumn{1}{c|}{10} & 9.79 & \multicolumn{1}{c|}{9} & 10.08 & 10.00 & -2.94 \\ \hline
90 & \multicolumn{1}{c|}{10} & 9.792 & \multicolumn{1}{c|}{9.5} & 10.12 & 5.00 & -3.35 \\ \hline
100 & \multicolumn{1}{c|}{10} & 9.792 & \multicolumn{1}{c|}{10} & 9.79 & 0.00 & 0.00 \\ \hline
\end{tabular}
% \begin{tablenotes}
% \begin{tiny}
% \item[\tnote{\textdagger}] 
% RT = Runtime (hours)
% \end{tiny}
% \end{tablenotes}
\end{table}

Table \ref{tab:case-study-time} shows the time reduction by SAF for the scaling done after different intervals of starting the simulation. 0\% and 100\% indicate that the entire simulation is run by the scaled and baseline architectures, respectively. For the multi-FPGA clusters, two cases are considered: (1) stopping execution \& starting over on scaled hardware, and (2) Skip scaling and running on baseline architecture only. The minimum time between these two is presented in the table. From the table, we can see that SAF can reduce the application runtime from 8.33\% to 25\% for this particular scenario. 

To calculate an estimate of energy consumption, we assume the static power of the FPGA and the average dynamic power of the application to be 22mW and 46mW, respectively. For SAF, we also considered the dynamic power (10mW) to maintain the reconfiguration of FPGAs used for scaling when they are waiting to be added to the architecture. We calculate the energy by adding the power consumptions of the FPGAs and multiplying it with the runtime and show in Table \ref{tab:case-study-time} that SAF can reduce up to 27.53\% of energy consumption compared to the clusters. SAF consumes slightly more energy for simulation completed 50\% or more since the multi-FPGA clusters use the baseline architecture with four FPGAs only.

\subsection{Low Resource Utilization}
In table \ref{tab:resource}, we show the resource utilization for PTRANS benchmarks compiled with the default shell and the SAF custom shell. From the table, we can see that the SAF custom shell utilizes a very small percentage (2\%) of additional resources, leaving plenty of resources for large-scale applications. 

\begin{table}[!hbt]
\scriptsize
\centering
\caption{\label{tab:resource} FPGA Resource Utilization for the default shell and SAF}
\begin{tabular}{|l|l|l|l|l|P{1.1cm}|}
\hline
\textbf{Shell} & \textbf{\begin{tabular}[c]{@{}l@{}}Logic\\ (ALMs)\end{tabular}} & \textbf{Reg} & \textbf{\begin{tabular}[c]{@{}l@{}}Block \\ Memory bits\end{tabular}} & \textbf{PLL} & \textbf{Pins} \\ \hline
Default & 43,994 (10\%) & 85,173 & 61,11,846 (11\%) & 60 (54\%) & 335 (41\%) \\ \hline
SAF & 49,861 (12\%) & 99,579 & 71,55,056 (13\%) & 60 (54\%) & 351 (43\%) \\ \hline
\end{tabular}
\end{table}

% \section{Scopes, Limitations and Future work}
%  The idea of connecting an accelerator to a remote host using standalone protocols can be applied to other accelerators for flexible scaling. Even though we implement and evaluate our framework using Intel Arria 10 FPGA and Intel FPGA SDK for OpenCL, a similar framework architecture can be implemented on FPGAs from other vendors having a shell with Ethernet support. A higher-level protocol like TCP/IP or an InfiniBand architecture \cite{infiniband} can be implemented on FPGA to extend the network capabilities of the framework. A single host managing multiple FPGA devices can be a performance bottleneck, as we can see from the performance scaling results. In that case, the number of FPGAs per host can be limited to a certain number (eight in our case). There might be packet loss during Ethernet transmission. There is no automatic retransmission logic for raw Ethernet protocol. In our work, we have inserted delays between packets depending on the packet size by trial and error. This is not ideal and can be improved by using a higher-level protocol like UDP or TCP/IP. Raw bitstream data is transmitted over the Ethernet network without any encryption used for bitstream security. In future work, encryption-decryption should be done on the host and device side, respectively, to ensure the secure transmission of bitstreams. 

\section{Conclusion}
In this paper, we present SAF, an Ethernet-based scalable acceleration framework for the flexible scaling of FPGAs. SAF provides a custom FPGA shell and a set of Ethernet protocols to operate FPGA in a standalone fashion without a local host. We introduce an automatic network discovery and remote configuration for multiple FPGAs that allow flexible scaling of FPGAs. The experiment results based on the Bittware 385A accelerator cards show faster configuration time, reduced setup cost, reduced runtime and energy consumption with an almost linear performance scaling. 

% \section*{Acknowledgment}

% % The preferred spelling of the word ``acknowledgment'' in America is without 
% % an ``e'' after the ``g''. Avoid the stilted expression ``one of us (R. B. 
% % G.) thanks $\ldots$''. Instead, try ``R. B. G. thanks$\ldots$''. Put sponsor 
% % acknowledgments in the unnumbered footnote on the first page.

% \avs{The capitalization in several of references is wrong.}
% For some reason, the bib template needs a double curly brace to retain the correct capitalization. I will correct it in the final version. 

%\avs{You need to fill the whole page with references.}

\newpage
\bibliographystyle{IEEEtran}
\bibliography{references.bib}
\end{document}